\documentclass[12pt]{article}

\usepackage{upgreek}
\usepackage{amsmath,amsfonts,amssymb}
\usepackage[pdftex]{graphicx}
\renewcommand{\figurename}{Fig.}

\begin{document}
\renewcommand{\figurename}{Fig.} 


\begin{center}
{\large{CHIRAL PHASE TRANSITION IN BARYON--DENSE MATTER WITH ACCOUNT
OF FINITE SIZE EFFECT}}\\
\bigskip
B.F. Kostenko$\,^{a,}$\footnote{E-mail:
bkostenko@jinr.ru},  J. Pribi\v{s}$\,^{b}$\\
\medskip
{\it{$^{a}$\small The Laboratory of Information Technologies, Joint Institute for Nuclear Research,\\ Dubna Russia}}\\ 
\medskip 
{\it{$^b$\small Faculty of Electrical Engineering and Informatics, Technical University in Ko\v sice, Slovakia}}\\
\end{center}

\begin{abstract}
We consider a model of chiral phase transitions in nuclei, suggested by T.D. Lee et
al., and argue that such transitions may be seen in cumulative effect investigations.
Finite-range effects arising from the smallness of the system are briefly discussed.
Some general proposals for future NICA and CBM experiments are given.
\end{abstract}

\section{Beyond parton description of cumulative effect}

\begin{figure}[!ht]\label{fig1}
\centerline{\includegraphics[width=11cm]{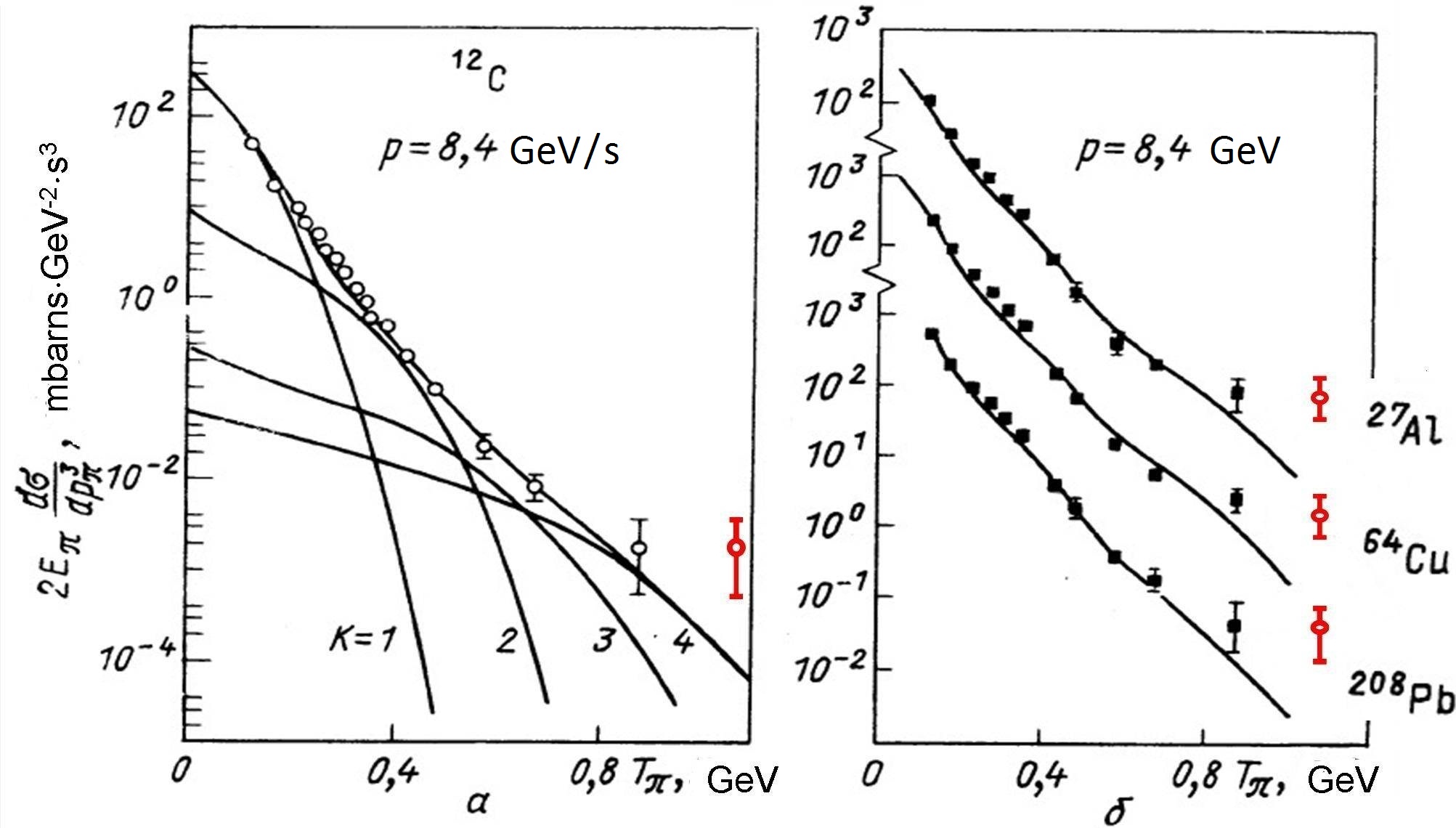}}
\vspace{-2mm}\caption{\small Cumulative $\pi$-meson production on $^{12}$C, $^{27}$Al, $^{64}$Cu, $^{208}$Pb. Contributions of fluctons containing  K= 2, 3 and 4 nucleons in $\pi^{+}$-meson yield are shown in Fig. 1.a.}
\end{figure}

In Fig.~\ref{fig1} experimental data~\cite{r01} on cumulative $\pi^+$-meson production in proton -- nucleus collisions at proton momentum $p=8.4$~GeV/c together with their theoretical description in the frame of the parton model~\cite{r02} are shown. One can see {\bf a strong deviation the theory and experiment} at $T_{\pi}=1.072$ GeV for all nuclei considered (the most discrepant experimental points are shown in red).

In Fig.~\ref{fig1}, one can see that $\pi^+$-meson yield becomes something like constant at $T_{\pi}> 0.9$~GeV.  This suggests {\bf a new mechanism} joins $\pi$-meson production in this region. A kinematical consideration shows that formation of a {\bf compound system} containing both the projectile proton and intranuclear K-nucleon system is the most simple and efficient mechanism for explanation the data. In particular, K = 3 is quite enough to explain the experiment~\cite{r01}. Interaction with K = 4 fluctons is nearly about of 20 times less probable, if one trusts the estimations of probabilities for 3- and 4-flucton appearance in nuclei given in~\cite{r02}.

\section{Signature of chiral phase transition in compound system}

Kinematics of cumulative particle production experiments permits to express a value of mass of K-baryon compound system emitted a cumulative meson via mass of (K$-$1)-baryon system colliding with the projectile.  In experiment~\cite{r01}, we have proton with momentum P bombarding a motionless 3-baryon system in initial state,
\medskip

\centerline{\includegraphics[width=8cm]{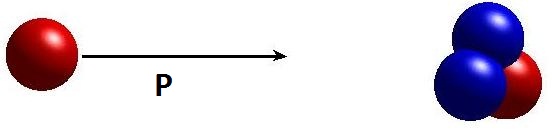}}
\medskip

\noindent and 4-baryon system emitted $\pi^+$-meson with momentum $-$(P$_1-$P) in final state,
\medskip

\centerline{\includegraphics[width=11cm]{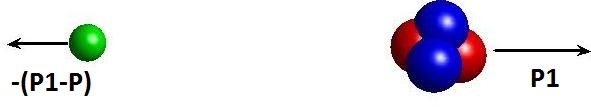}}
\medskip

\noindent then the energy conservation law gives: E$_p$ + M$_3$ = E$_{\pi}$ + (P$_1^2$ + M$_4^2$)$^{1/2}$, where E$_p$ = (P$^2$ + M$_p^2$)$^{1/2}$,   E$_{\pi}$ =((P$_1$ - P)$^2$ + M$_{\pi}^2$)$^{1/2}$, and P,  (P$_1 -$P) are known from the experimental conditions.  This allows to express the mean energy per baryon in 4-baryon systems, m$_4$ = M$_4$/4, through that of 3-baryon system, m$_3$ = M$_3$/3, and calculate the difference m$_4-$m$_3$.  Leaving aside for a while some subtleties, one may assume that the condition m$_4 -$m$_3< 0$ is a possible signature of the chiral phase transition.

Numerical estimation of (m$_4 -$m$_3$)-dependence on m$_3$ for cumulative $\pi^+$-meson production in experiment~\cite{r02} is shown in Fig.2. 
\begin{figure}[!ht]\label{fig2}
\centerline{\includegraphics[width=12cm]{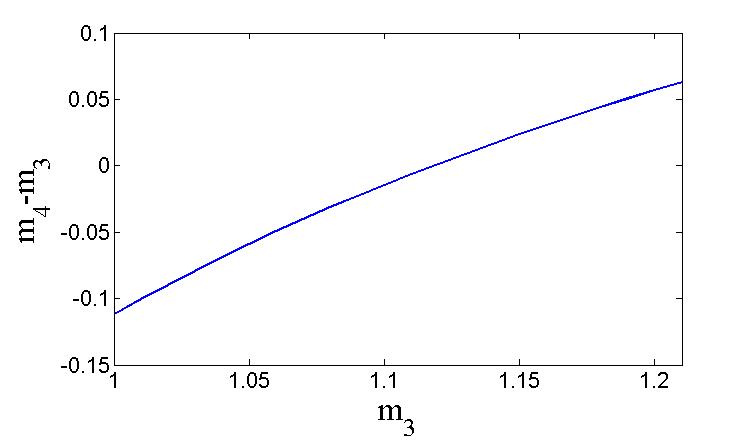}}
\vspace{-2mm}\caption{\small Change of energy per baryon in compound 4-baryon system after emission of cumulative $\pi^+$-meson as compared with energy per baryon in 3-baryon flucton preexisting in nuclei before interaction. The condition m$_4 -$m$_3 < 0$ is a possible signature of the chiral phase transition in experiment~\cite{r01}.}
\end{figure}

\section{Some theory}

There is a big uncertainty in our knowledge of flucton characteristics. Different quark bag models~\cite{r02,r03,r04} give different values for M$_3$, M$_4$ and M$_5$, though they all predict the minimal experimentally observed mass of dibaryon resonance, M$_2=2.15$~GeV/c$^3$~\cite{r05}. Paper~\cite{r03} predicts that m$_3\approx 1.05$ and ${\rm m}_3- {\rm m}_4 < 0$, papers~\cite{r02} and~\cite{r04} give m$_3\approx 1.2$ and m$_3\approx 1.17$, accordingly, and 
${\rm m}_3- {\rm m}_4 > 0$.  But all of them suggest that chiral phase transition {\bf already happens} in the flucton before its interaction with projectile, since masses of u- and d- quarks is supposed to be close to zero. If it is so, then the mass change possibly observed in~\cite{r01} is a sequence of change of color-electric and color-magnetic interaction between quarks (QCD predicts their decrease in the dense baryon matter).

However, it should be noted that a point of view that the flucton is really a dense multinucleon droplet with zero lifetime {\bf prevails now}. It is the model of shot-range correlations. This suggestion was recently confirmed in electron scattering experiments~\cite{r06}. The model predicts a continuous spectrum of masses starting from m$_3=1$. In this case scattering by 3-baryon systems with 
m$_3< 1.118$ gives an {\bf evidence for chiral phase transition} (see Fig. 2).  
To establish this, scattering by heavier 3-baryon system should be experimentally separated (see below).  

Quark bag models can be certainly used for description of the final state of the compound system after the cumulative particle ejection, if phase transition from nuclear matter into another phase actually takes place. One can see in Fig. 2 that $({\rm m}_4 - {\rm m}_3)/{\rm m}_3 \ll 1$. The closeness of m$_4$ to m$_3$ (which is the mean energy per baryon in 3-baryon system {\bf belonging to the ground state of nuclei}) indicates existence of {\bf strong quenching} of the 4-baryon compound system due to the cumulative meson emission. Because of this, for estimation of different characteristics of the final state we can use models explaining the mass spectrum of elementary particles and multibaryon ground states. Proceeding along this way, we obtain: 

\begin{itemize}
\item {\bf Shell effects.} Models [2 -- 4] give approximate constant value of m$_K$ for K$=$3, 4, 5. One can expect only a small (on 2-3 \% accuracy level) deviation from the "bulk"$\,$ consideration shown in Fig. 2. 
\item {\bf Surface tension.} According to~\cite{r03} the surface tension coefficient $\sigma\leq (70 {\rm{MeV}})^3$ $\approx 8.8$ MeV/fm$^2$. This gives for m3 a correction on 2-3 \% level for radius of the flucton  R$\approx0.8$~fm~\cite{r02}. Independent consideration of the Casimir energy in the Chiral bag model framework~\cite{r07} gives the same estimation. 
\item {\bf Coulomb energy.}  A compound (N+1)-baryon system, consisting of the projec\-tile proton and the intranuclear N-baryon system, may obtain an additional mass increase due to the Coulomb repulsion of the charge of projec\-tile with a charge of the N-baryon system. This gives a correction to m$_3$ on 0.13 \% level or less, which is a small value too.
\end{itemize}

\section{Predictions and proposal}

Using previous consideration, it is possible to predict results of a cumulative production experiment through creation of 5-baryon compound system. Such results could be observed in the experiment~\cite{r01}, if number of observed events were increased in 20 times (we use estimation for the probability of existence of 4-baryon fluctons according to~\cite{r01}).
\begin{figure}[!ht]\label{fig3}
\centerline{\includegraphics[width=12cm]{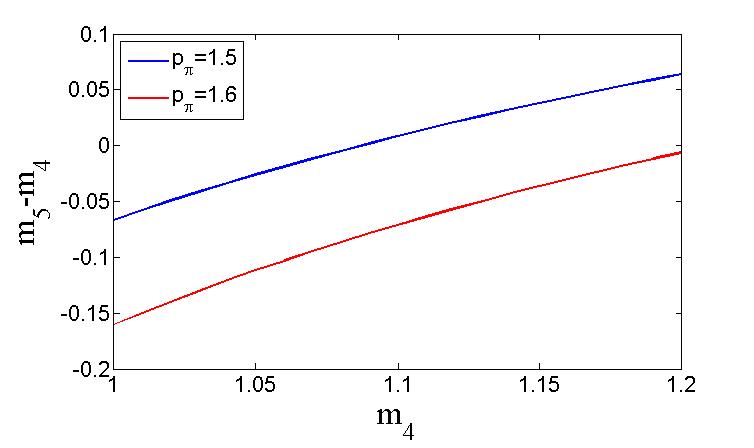}}
\vspace{-2mm}\caption{\small This figure predicts a possibility to observe cumulative $\pi^+$-mesons with momentum P$_{\pi}\approx 1.5$ GeV/c in experiment of Baldin's type~\cite{r01}.}
\end{figure}
Indeed, one can expect that dependence of the difference $({\rm m}_5 - {\rm m}_4)$ on m$_4$ should be close to $({\rm m}_4 - {\rm m}_3)$-on-m$_3$ dependence, since the effect of mass change is of a bulk type and not expected to depend strongly on the mean baryon mass in the flucton. Energy release caused by chiral phase transition leads to an extra increase of cumulative meson energy. It is convenient to tag this "explosive situation"$\,$ with a special term: {\bf chiral shot (or chiral discharge)}. Formation of the compound system and the chiral discharge may be both responsible for essential increase of energy of cumulative particles. 

We also propose to study proton -- light ion (from He to C) collisions with high momentum cumulative $\pi^+$-meson trigger.  Registration of secondary nucleons, which have a considerable value to leave a light target nucleus, allows us to determine m4 directly. Integration these results with experimental estimations of m3 (see~\cite{r06} as a prototype) allows to obtain direct information about dependence of $({\rm m}_4 - {\rm m}_3)$ on m$_3$. The dependence of $({\rm m}_5 - {\rm m}_4)$ on m$_4$ may be also feasible.

Investigation of the collisions mentioned above with other triggers is also interesting. Proton triggers allows studying phase transitions without change of baryon number of the compound system. K$^+$-meson trigger gives information on strange quark presence in the excited compound system. 

In this respect it is interesting to note that difficulties with the description of the strange particles within the parton model framework were pointed out as far back as 1982~\cite{r08}.  If a model assumed here is true, then characteristics of the compound system in the experiment~\cite{r01} might be so high (temperature up to 1.25 GeV, $\rho$ up to $10\, \rho_0$, where $\rho_0 = 0.17$~fm$^{-3}$ is baryon density in nuclei), that the system certainly underwent a violent phase transition into quark-gluon plasma. At these conditions the strange quark mass becomes close to its bare (current-algebra) value, m$_s \leq 100$~MeV~\cite{r09}, and production of strange mesons should sharply increase. Experiments on cumulative production of K$^+$ mesons confirm the conclusion~\cite{r10}. Therefore, it looks quite possible that {\bf registration of excited strange matter}, the ground state of which is believed to exist inside neutron stars, were accomplished for the first time in experiments on cumulative production of K$^+$-mesons.

Investigation of doubly cumulative processes that can take place in flucton-flucton collisions gives a unique possibility to create compound systems in highly excited and the ground states with density even higher than density of an individual flucton [10 -- 12].  For example, it is possible to create an initial state with two 4-baryon fluctons 

\centerline{\includegraphics[width=11cm]{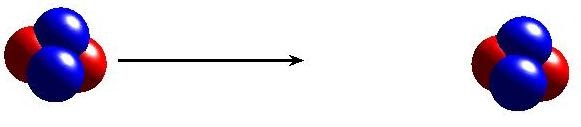}}

\noindent and a final state with cumulative $\pi$-meson and 8-baryon flucton
\medskip

\centerline{\includegraphics[width=12cm]{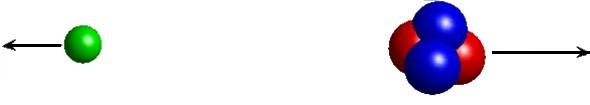}}

\noindent Such processes provide densities $\rho = 20\, \rho_0$ and temperatures up to kinetic energy of the projectile 4-baryon flucton divided by a baryon number of the compound system for an excited compound state and till T$\,=0$ for the ground state. It is clear that {\bf none of existing or planned up to now experiments can ensure such conditions}. Collisions of fluctons with higher baryon numbers may lead not only to creation of strange matter, but also to charmed one~\cite{r11}.

\section{Instead of conclusion}

Phase transition in Quantum Field Theory is understood as vacuum transformation and, therefore, may be attributed even to a system without physical particles, to say nothing of system with finite number of them. Nevertheless, if one says that a high value of baryon density may induce phase transition this means that baryon density is an approximate "macroscopic"$\,$ parameter. The baryon number conservation law plays the decisive role in that interpretation. Here, we have {\bf assumed} that the notion of phase transition is already applicable, at least in some broad-brush picture, to systems enclosing 4 -- 8 baryons.
A chiral phase transition model at high baryon densities in bulk matter in which nucleon loses all or a part of its mass was developed by T.D. Lee et al.~\cite{r13}.  A more recent paper~\cite{r14} suitable also for baryon droplets was published by M. Alford et al.~\cite{r14}. 


\end{document}